\documentclass[journal=arxiv,manuscript=article]{achemso}
\usepackage{color}

\usepackage[version=3]{mhchem} 

\author{Raphael M. Tromer}
\affiliation[State University of Campinas]
{Applied Physics Department, State University of Campinas, Campinas, SP, 13083-970, Brazil}
\alsoaffiliation[State University of Campinas]
{Center for Computational Engineering and Sciences, State University of Campinas, Campinas, SP, 13083-970, Brazil}
\author{Levi C. Felix}
\alsoaffiliation[State University of Campinas]
{Center for Computational Engineering and Sciences, State University of Campinas, Campinas, SP, 13083-970, Brazil}
\affiliation[State University of Campinas]
{Applied Physics Department, State University of Campinas, Campinas, SP, 13083-970, Brazil}
\author{Luiz A. Ribeiro}
\affiliation[University of Brasilia]
{Institute of Physics, University of Bras\'ilia , 70919-970, Bras\'ilia, Brazil}
\author{Douglas S. Galvao}
\alsoaffiliation[State University of Campinas]
{Center for Computational Engineering and Sciences, State University of Campinas, Campinas, SP, 13083-970, Brazil}
\email{galvao@ifi.unicamp.br}
\affiliation[State University of Campinas]
{Applied Physics Department, State University of Campinas, Campinas, SP, 13083-970, Brazil}

\title{Optoelectronic Properties of Amorphous Carbon-Based Nanotube and Nanoscroll}

\begin{document}


\begin{abstract}
Free-standing monolayer amorphous carbon (MAC) is a pure carbon structure composed of randomly distributed atom rings with different sizes, which was recently synthesized. In this work, we carried out ab initio and tight-binding calculations to investigate the optoelectronic properties of MAC and its derived nanotube and nanoscroll configurations. Our results show MAC, tube, and scrolls exhibit similar electronic behavior. All structures absorb from infrared to ultraviolet, with maximum absorption peaks the visible-ultra violet ($\sim 3.2$ eV). The maximum and minimum reflectivity values are in the range 0.3-0.5 (infrared) and 0.1-0.0 (ultraviolet), making these materials good candidates to ultraviolet filters.
\end{abstract}


\section{Introduction}
Carbon-based nanomaterials are currently among the best candidates for obtaining new green-energy solutions with a good cost-efficiency ratio \cite{li_EC_2020,tsang_RSER_2020,lux_CEC_2015}. Among these solutions, organic photovoltaics \cite{ryu_NS_2020} and light-emitting diodes \cite{zhang_CSR_2020} have already reached the market by showing considerable potential to overcome silicon-based electronics as the next state-of-the-art in such kinds of applications. The rapid growth in obtaining graphene-based prototype devices stimulated several theoretical and experimental investigations to obtain new structures to create more efficient materials \cite{li_EC_2020,tsang_RSER_2020}. Very recently, the synthesis of the first free-standing monolayer of amorphous carbon (MAC) was obtained by Toh and coworkers \cite{toh_2020}. MAC is an amorphous material with sp$^2$ and sp$^3$-like hybridized carbon atoms arranged in a lattice with a wide distribution of bond length and bond angle values and containing five- to up eight-member rings \cite{toh_2020}. It is worthwhile to stress that short-range order exists in MAC, but the inter-atomic distances and inter-bonding angles are different regarding pristine graphene lattices.

The physical/chemical properties of amorphous carbon materials have been extensively investigated in the past few years for proposing new routes for carbon use in nanoelectronics \cite{falcao_JCTB,joo_SciAdv,toh_2020,kotakoski_2011,eder_2014,mortazavi_2016,rahaman_2017}. For instance, some theoretical works have investigated the mechanical properties of an amorphized graphene lattice generated by randomly distributing pentagonal and heptagonal rings \cite{mortazavi_2016,rahaman_2017}. Such a lattice possesses a reasonable yield strength when compared to other two-dimensional carbon allotropes. In experimental studies, a Zachariasen carbon monolayer --- a new amorphous 2D carbon allotrope with one-atom-thick --- was synthesized by Joo and colleagues \cite{joo_SciAdv}. In their work, an in-plane fully sp2-like hybridized carbon network was obtained at high temperatures ($>900^{\circ}$C). As mentioned above, very recently, laser-assisted chemical vapor deposition was employed to synthesize a stable MAC \cite{toh_2020}. MAC can be deformed at high load strain values without breaking and without crack propagation. Importantly, the synthesized MAC has a ring distribution, which makes MAC a unique new carbon nanostructure.  

In this work, motivated by the recent MAC synthesis \cite{toh_2020}, the electronic and optical properties of MAC lattices in the form of sheet, nanotube, and nanoscroll were investigated in the framework of density functional theory (DFT) and DFT-tight-binding (DFTB) calculations. The present study addresses the effects of structural curvature on the electronic properties (bandgap values, charge localization, and photon absorption behavior) of MAC-based structures.     

\section{Methods}
The systems considered here (see Figure \ref{fig:structures}) have a large number of atoms (610 carbon atoms), to be fully investigated using ab initio methods. To address this limitation, we have adopted the combined use of DFTB (geometry optimization and electronic structure) and DFT (photon absorption calculations, using the optimized geometries from DFTB simulations. 

For the DFTB calculations, we used a self-consistent charge density functional tight-binding (scc-dftb)) method, as implemented in the DFTB+ software \cite{dftb}. DFTB+ is derived from ab initio calculations based on a second-order expansion of the total energy. Therefore, it can treat large systems in a fast and accurate fashion. The photon absorption calculations are not possible with DFTB+. 

For the DFT calculations we used a linear scaling ab initio method that is based on localized atomic orbitals and numerical atomic orbital definition, as implemented in SIESTA software \cite{Soler_2002}. Through the use of a linear combination of these numerical atomic orbitals, SIESTA can efficiently perform calculations with considerable accuracy and reduced computational cost (in relation to other plane waves methods, such as QUANTUM ESPRESSO \cite{Giannozzi2009}. 

 We have investigated the electronic and optical properties of MAC based  structures in three distinct configurations: sheet, tube, and scroll (see Figure \ref{fig:structures}). They contain the same number of atoms (610) in their unit cell.
 
 For the MAC sheet, an orthorhombic unit cell was considered with lattice vectors given by $l_x=l_y=40.0$ \AA~and $l_z=25$ \AA (to introduce a vacuum region), which are the same parameters from the original MAC paper \cite{toh_2020}. The carbon atom positions are fully optimized. The other structures were generated by rolling up the MAC sheet along the z-direction, forming a tube, and two species of scrolls with a different internal radius (see Figure \ref{fig:structures}).  

The used DFTB+ set parameter was the pbc-0-3 \cite{pbc-03}. This parameter set consists of Slater-Koster coefficients especially developed to treat solid-state systems. van der Walls effects were included in the relaxation process. The convergence criterion of $0.05$ eV/\AA~for the force was used during geometry optimization and the convergence threshold on the self-consistent charge (SCC) procedure was set as $0.0001$ in $e$ units. For the orthorhombic cell, the k points were generated within the Monkhorst-Pack scheme using a $4\times4\times1$ mesh for sheet and $1\times 1\times 4$ for tube and scroll.

As we have mentioned above, SIESTA was used here only to perform photon absorption calculations. These calculations were carried out within the generalized gradient approximation (GGA) with Perdew-Burke-Ernzenhof (PBE) exchange-correlation functional \cite{Perdew_1996}. Norm-conserving Troullier-Martins pseudo-potentials were used to describe the core electrons, and a double-zeta plus polarization (DZP) basis set for representing the wave functions. The convergence of the kinetic energy cut-off was achieved with $150$ eV, and we sampled the reciprocal space using only $\Gamma$ point.

The optical properties are obtained by considering the complex dielectric function expression $\epsilon =\epsilon_1+i\epsilon_2$. The imaginary part, $\epsilon_2$, can be obtained from Fermi's golden rule \cite{Fermi},
\begin{equation}
\epsilon_2(\omega)=\frac{4\pi^2}{\Omega\omega^2}\displaystyle\sum_{i\in \mathrm{VB},j\in \mathrm{CB}}\displaystyle\sum_{k}W_k|\rho_{ij}|^2\delta	(\epsilon_{kj}-\epsilon_{ki}-\omega),
\label{eq1}
\end{equation}
where VB and CB denote the valence and conduction bands respectively, $\rho_{ij}$ is the dipole transition matrix element, $\omega$ is the frequency of the photon, $W_K$ the k points weight and $\Omega$ is the unit cell volume. The real and imaginary parts of the dielectric function $\epsilon_1$ and $\epsilon_2$ are related through the Kramers-Kronig relation \cite{Kramers-Kronig}, in which $\epsilon_1$ is expressed by:
\begin{equation}
\epsilon_1(\omega)=1+\frac{1}{\pi}P\displaystyle\int_{0}^{\infty}d\omega'\frac{\omega'\epsilon_2(\omega')}{\omega'^2-\omega^2},
\label{eq2}
\end{equation}
where $P$ denotes the principal value. All other optical quantities, such as the absorption coefficient $\alpha$, reflectivity $R$, refractive index $\eta$, and loss function $L$, can be readily calculated once $\epsilon_1$ and $\epsilon_2$ are determined:
\begin{equation}
\alpha (\omega )=\sqrt{2}\omega\bigg[(\epsilon_1^2(\omega)+\epsilon_2^2(\omega))^{1/2}-\epsilon_1(\omega)\bigg ]^{1/2},
\end{equation}
\begin{equation}
R(\omega)=\bigg [\frac{(\epsilon_1(\omega)+i\epsilon_2(\omega))^{1/2}-1}{(\epsilon_1(\omega)+i\epsilon_2(\omega))^{1/2}+1}\bigg ]^2 ,
\end{equation}
\begin{equation}
\eta(\omega)= \frac{1}{\sqrt{2}} \bigg [(\epsilon_1^2(\omega)+\epsilon_2^2(\omega))^{1/2}+\epsilon_1(\omega)\bigg ]^{2},
\end{equation}
and
\begin{equation}
    L(\omega)=\frac{\epsilon_2(\omega)}{(\epsilon_1(\omega)^2+\epsilon_2(\omega)^2).}
\end{equation}

\section{Results}

\subsection{Electronic Analysis} 
\begin{figure}[htbp!]
\begin{center}
\includegraphics[width=0.9\linewidth]{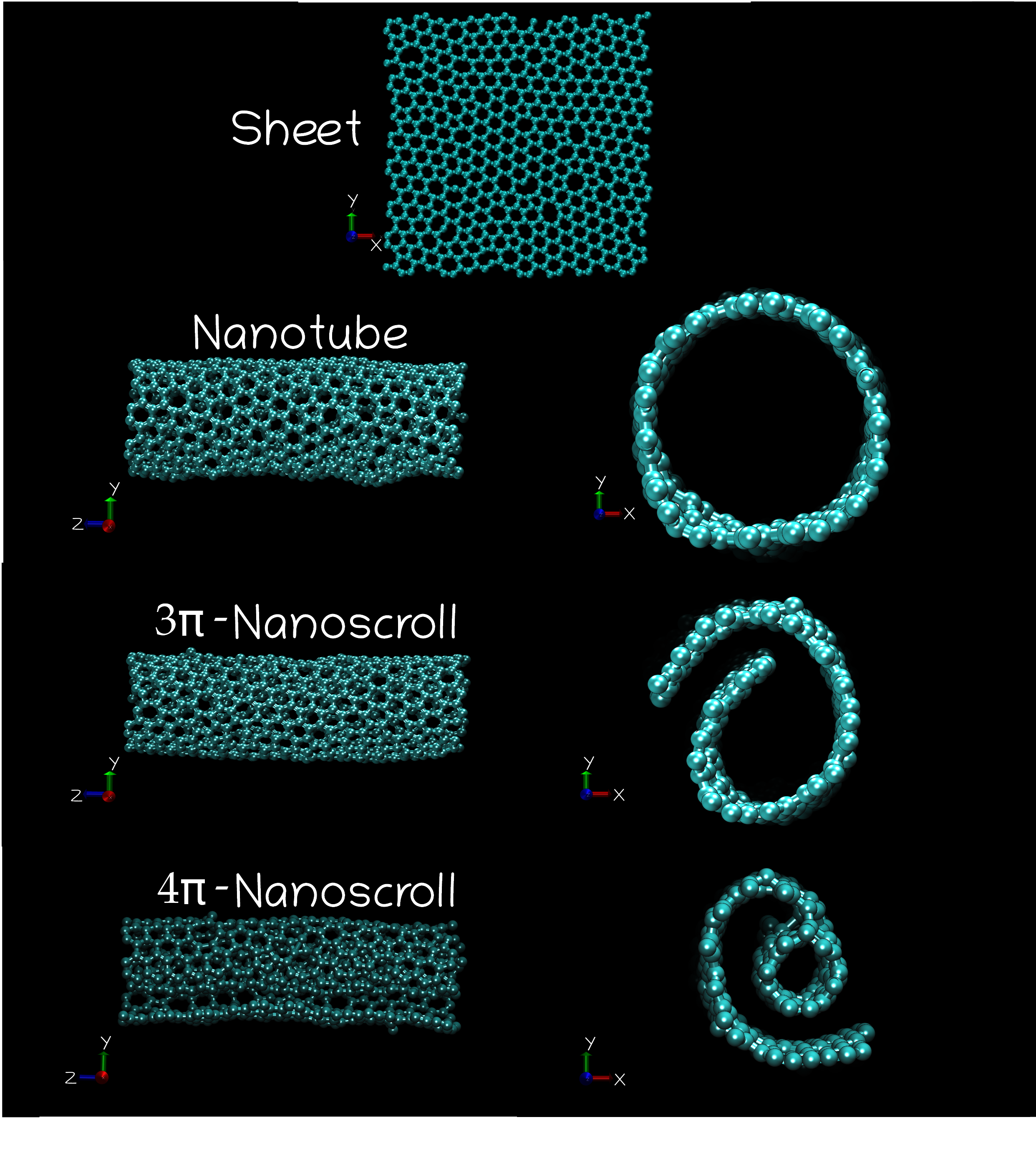}
\caption{Optmized MAC-based structures from DFTB+ calculations.}
\label{fig:structures}
\end{center}
\end{figure}

In Figure \ref{fig:structures} we present the optimized structures obtained from the DFTB+ calculations. For the MAC sheet case, our optimization procedure yields a maximum difference between carbon-carbon bonds about $0.1$~\AA. The average values for the carbon-carbon bonds are $1.414$~\AA, $1.431$~\AA, $1.434$~\AA, and $1.437$~\AA~for the MAC sheet, nanotube, $3\pi$-nanoscroll, and $4\pi$-nanoscroll, respectively. We observed that the bond lengths obtained for curved structures are more stretched when compared to the MAC sheet. In this sense, for the structures considered here, the curvature effects do not substantially affect the bond length values. It is worthwhile to stress that for the scroll structures there are no covalent bonds between layers.

\begin{figure}[htbp!]
\begin{center}
\includegraphics[width=1.1\linewidth]{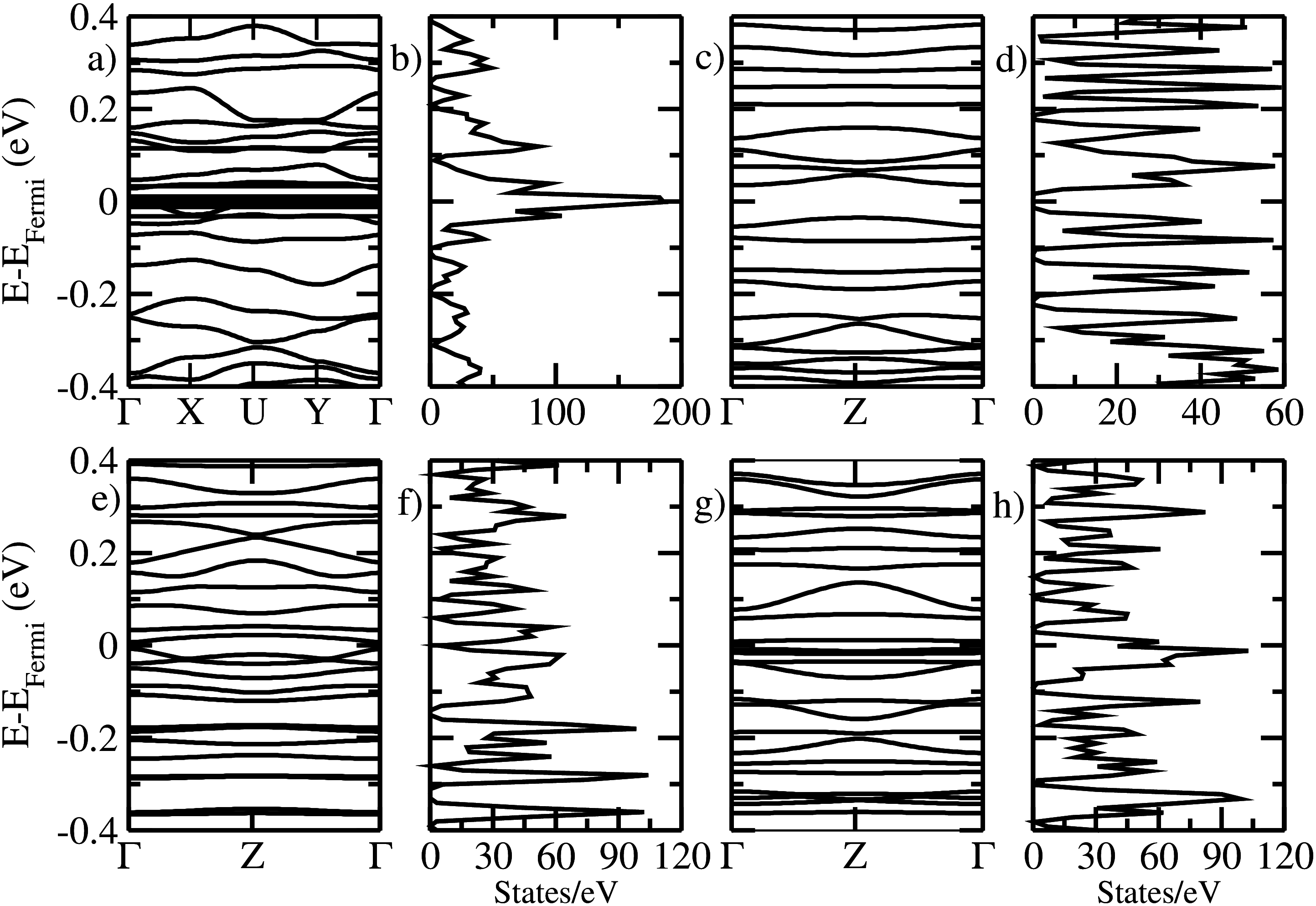}
\caption{Electronic band structure and the corresponding total density of states (DOS) for MAC-based (a-b) sheet, (c-d) nanotube, (e-f) $3\pi$-nanoscroll, and (g-h) $4\pi$-nanoscroll.}
\label{fig:bands}
\end{center}
\end{figure}

In Figure \ref{fig:bands} we present the band structures and the corresponding total density of states (DOS) for the structures presented in Figure \ref{fig:structures}. MAC sheet (Figures \ref{fig:bands}(a-b)) presents a peak at the Fermi level (zero-point in the panels of Figure \ref{fig:bands}), which indicates a metallic behavior. These results are in agreement with the ones reported by Toh et al. \cite{toh_2020}. 

For the MAC-based nanotube (Figure \ref{fig:bands}(c-d)), $3\pi$-nanoscroll (Figure \ref{fig:bands}(e-f)), and $4\pi$-nanoscroll (Figure \ref{fig:bands}(g-h)), the band structures and DOS present small bandgap values ($90$, $13$, and $18$ meV, respectively). These values are smaller than the  thermal energy at room temperature, which is $25$ meV. Thus, although the curvature effects tend to induce bandgap openings, the electronic behavior remains essentially metallic. The observed flatness of some bands is expected due to the amorphous (high disorder) character of the structures (the Dirac cone is not present in the MAC sheet).

\begin{figure}[htbp!]
\begin{center}
\includegraphics[width=1.0\linewidth]{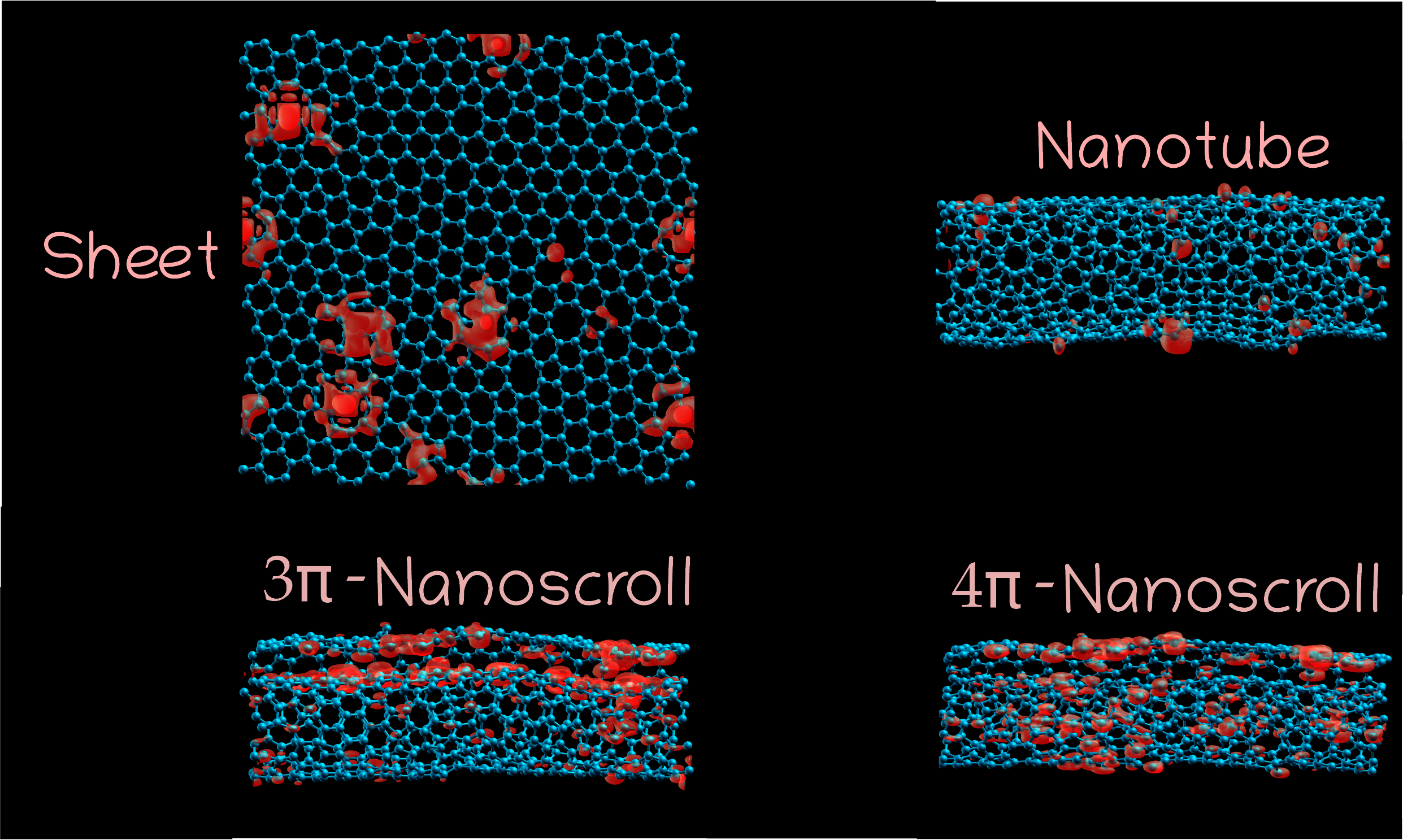}
\caption{Schematic representation of the lowest occupied crystal orbital (LUCO) for the structures shown in Figure \ref{fig:structures}.}
\label{fig:luco}
\end{center}
\end{figure}

In Figure \ref{fig:luco}, we present the spatial distribution of the lowest unoccupied crystal orbital (LUCO). In relation to the sheet and tube, as expected, the scrolls present a larger charge localization at the borders, due to the dangling bonds. The highest occupied crystal orbital (HOCO) patterns are similar.

\subsection{Optical Absorption Analysis}

The optical absorption analysis were performed considering a polarized light averaged over the x, y, and z directions. The calculated optical spectra for the real and imaginary parts of the dielectric function are displayed in Figure S1 in the Supplementary Material. For the calculations performed here, the considered photon spectral energy range was $0-5$ eV. The static dielectric constant $\epsilon_1(0)$ depends estimated from Figure S1(a) are $\epsilon_1(0)^{sheet}=17.6$, $\epsilon_1(0)^{nanotube}=12.6$,  $\epsilon_1(0)^{3\pi-nanoscroll}=43.78$ and $\epsilon_1(0)^{4\pi-nanoscroll}=27.7$. For comparison, the value for the graphene sheet is $\epsilon_1(0)^{graphene}=7.7$ \cite{Kumar_2019}. The largest values for the scrolls can be attributed to their larger discontinuous charge localization, as illustrated in Figure \ref{fig:luco}.

\begin{figure}[htbp!]
\begin{center}
\includegraphics[width=0.9\linewidth]{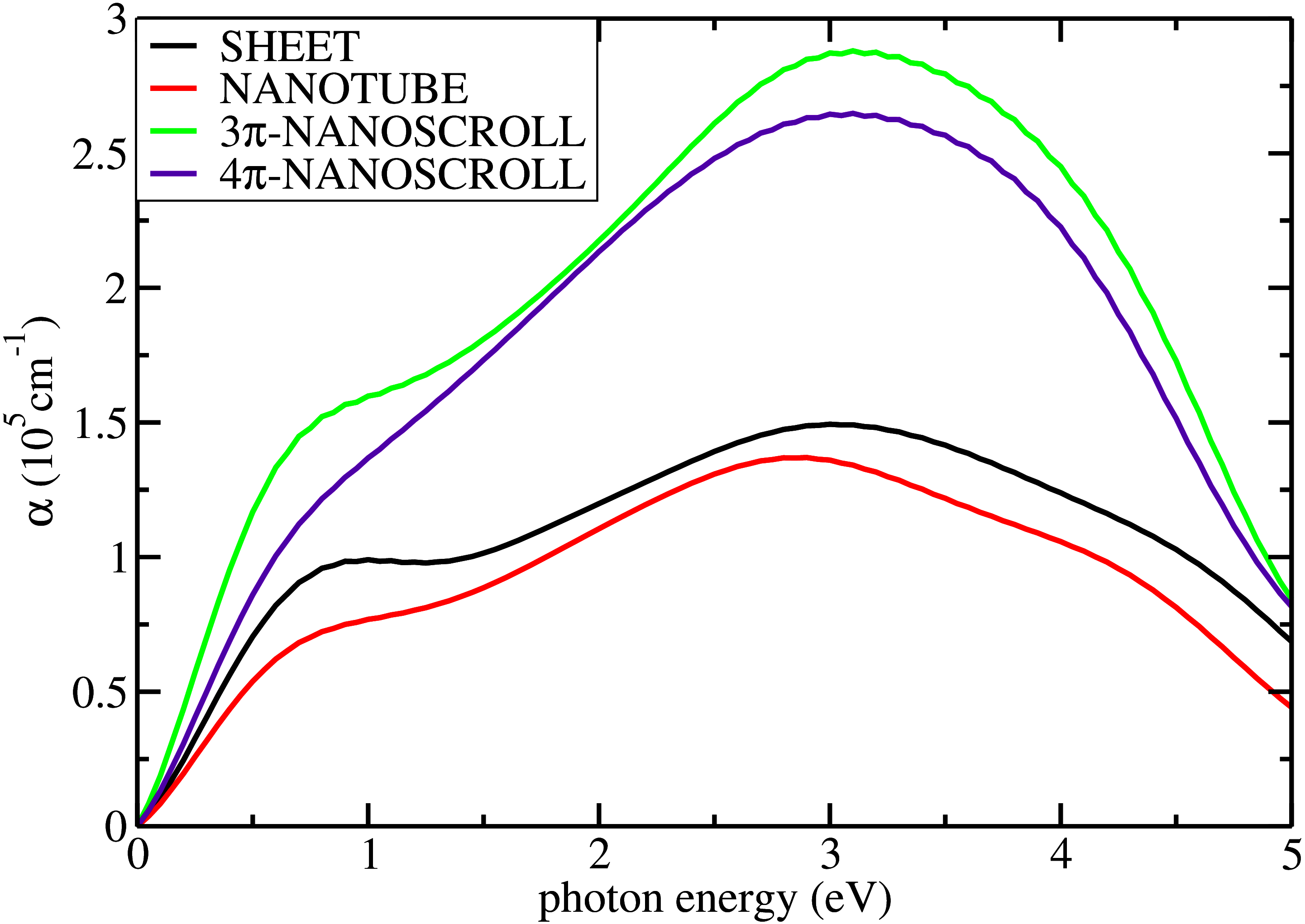}
\caption{Absorption coefficient as function of photon energy values. }
\label{fig:absorption}
\end{center}
\end{figure}

In Figure \ref{fig:absorption} we present the absorption coefficient as a function of photon energy for the structures shown in Figure \ref{fig:structures}. The first peak occurs when photon energy is less than $1$ eV, in the infrared region, for all cases. The maximum absorption peaks are localized in the visible-UV edge when the photon energy is approximately $3.2$ eV. We can see that the structures show a similar absorption pattern (in terms of energy range), with the main differences being the intensity values. The large curvatures of the nanoscroll increase the intensity of light absorption of almost two times in the ultra-violet region in relation to their corresponding tube.

\begin{figure}[htbp!]
\begin{center}
\includegraphics[width=0.9\linewidth]{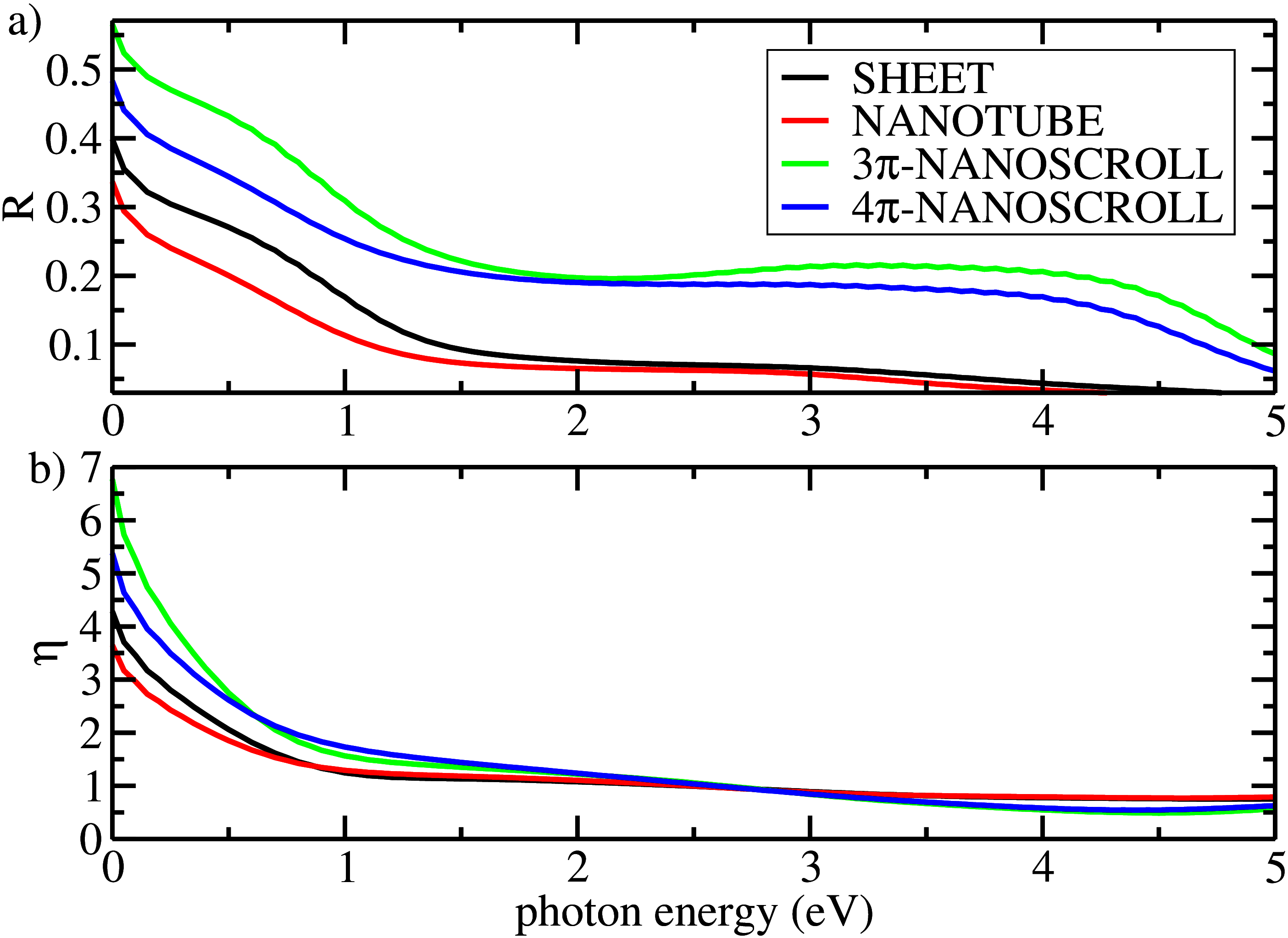}
\caption{(a) Reflectivity $R$ and (b) Refractive index $\eta$ as function of photon energy values. }
\label{fig:ref}
\end{center}
\end{figure}

The possibility of reflecting light is also an important feature that should be discussed. In Figure \ref{fig:ref} we present the reflectivity $R$ and refractive index $\eta$ as a function of photon energy. The maximum value of reflectivity is observed when photon energy is zero, as expected from their metallic behavior. In this regime, $-3\pi$-nanoscroll exhibits the highest $R$ value. The reflectivity is maximum in the infrared range. When photon energy increases up to $2$ eV, the reflectivity $R$ falls to $0.2$ for the nanoscroll cases and $0.1$ for the sheet and nanotube. This behavior continues to higher energy values, about 4 eV in the ultraviolet range. The reflectivity tends to small values after that, as shown in Figure \ref{fig:ref}(a). 

Comparing to Figure \ref{fig:ref}(b) when refractive index tends to 1.0 eV at high energies, we conclude that the material is not transparent and the major part of incident light will be absorbed, indicating that structures can be used in optoelectronics for applications as ultraviolet filters. From Figure \ref{fig:ref}(b), we can extract static refractive index $\eta (0)$ obtained for photon energy equal to zero. The obtained values are: $\eta (0)^{sheet}=4.26$, $\eta (0)^{nanotube}=3.61$, $\eta (0)^{3\pi-nanoscroll}=6.75$, and $\eta (0)^{4\pi-nanoscroll}=5.33$. The graphene value is $\eta (0)^{graphene}=2.53$, as reported in references \cite{optic1,optic2}.

\section{Conclusions}

In summary, we carried out DFT and DFTB+ calculations to investigate the electronic and optical properties of a monolayer of amorphous carbon into three different configurations: sheet, nanotube, and nanoscroll. 

Our findings revealed that curvature effects do not produce substantial differences in the electronic properties when contrasted with the results obtained for the planar structure. While the sheet is metallic, the curved structures present a small bandgap opening with values of $90$, $13$, and $18$ meV for nanotube, $3\pi$-nanoscroll, and $4\pi$-nanoscroll, respectively. 

For optical properties, the structures absorb from infrared to ultraviolet. The maximum absorption peaks are localized in the visible-UV ($\sim 3.2$ eV). The structures show a similar absorption pattern (in terms of energy range), with the main differences being the intensity values, the scrolls absorb two times more in the ultra-violet region than the tube.

The maximum and minimum reflectivity values are in the range 0.3-0.5 (infrared) and 0.1-0.0 (ultraviolet), making these materials good candidates to ultraviolet filters. The maximum and minimum reflectivity are observed for the scrolls and sheet/tube, respectively.

Considering that the MAC sheet has already been synthesized \cite{toh_2020} and their scroll fabrication is within the reach of our present technology \cite{scroll} (the tube would be a more challenging task), we hope the present work will stimulate further studies on this new carbon allotrope structure. 

\begin{acknowledgement}
The authors gratefully acknowledge the financial support from Brazilian research agencies CNPq, FAPESP, and FAP-DF and CENAPAD-SP for providing the computational facilities. L.A.R.J acknowledges the financial support from a Brazilian Research Council FAP-DF and CNPq grants $00193.0000248/2019-32$ and $302236/2018-0$, respectively. DSG thanks the Center for Computing in Engineering and Sciences at Unicamp for financial support through the FAPESP/CEPID Grants \#2013/08293-7 and \#2018/11352-7.
\end{acknowledgement}
                                                 


\begin{suppinfo}
The Supplementary Material contains the figures for the real and imaginary parts of dielectric constant function (Equations \ref{eq1} and \ref{eq2}, respectively), as well as a figure for the interplay between the loss function and the photon energy to determine the energy associated with plasma frequency $\omega_p$, namely plasma energy ($\hbar\omega_p$).
 
\begin{figure}[htbp!]
\begin{center}
\includegraphics[width=0.9\linewidth]{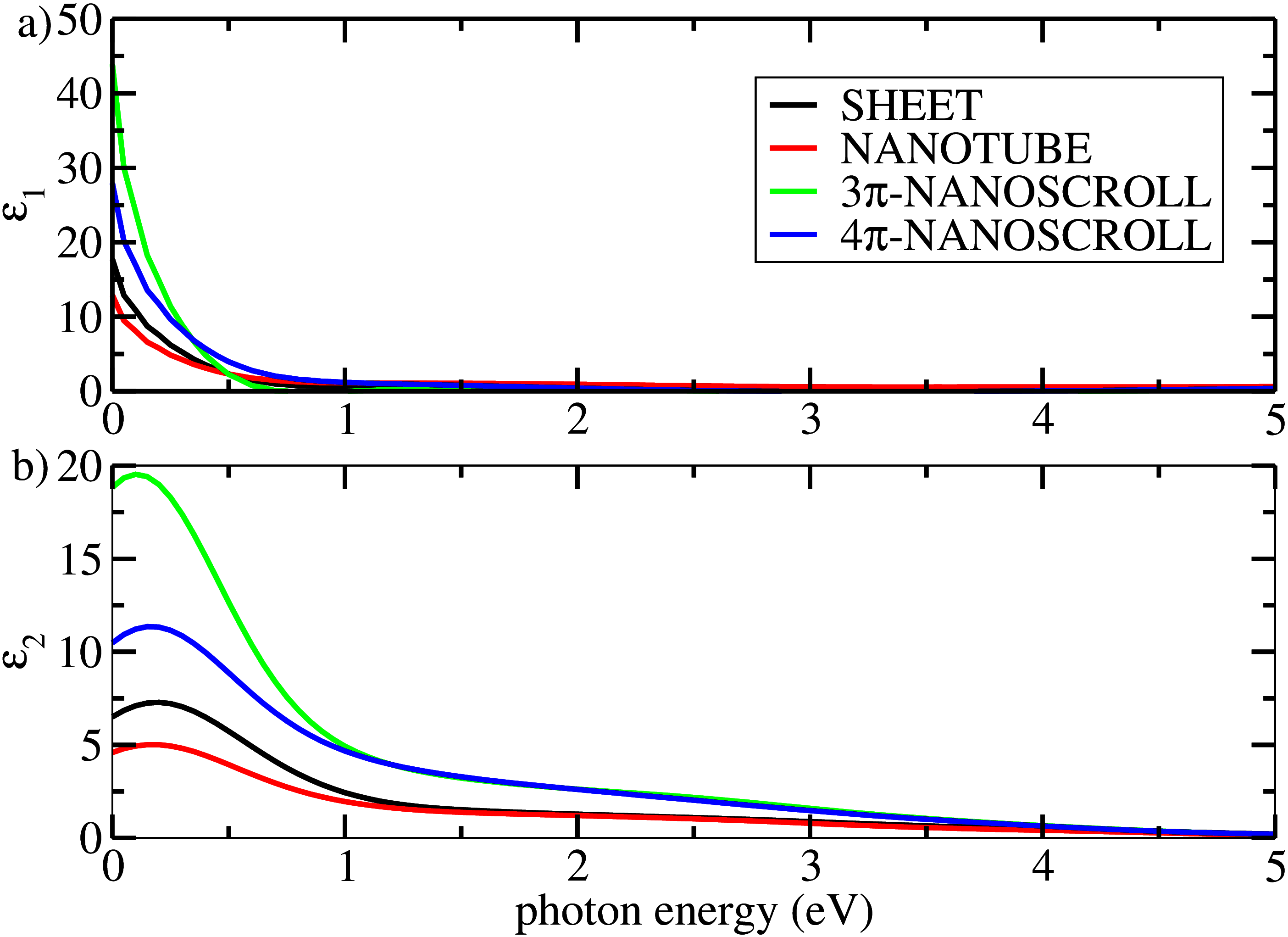}
\caption{(a) Real and (b) imaginary parts of dielectric constant function (Equations \ref{eq1} and \ref{eq2}, respectively). In Figure \ref{fig:epsilon}(b), we can see that the first peak is located for photon energy equals to zero for all structures, which is a consequence of their metallic behavior. As small red shift (in relation to the sheet) is observed for the tube and scrolls.}
\label{fig:epsilon}
\end{center}
\end{figure}

\begin{figure}[htbp!]
\begin{center}
\includegraphics[width=1.0\linewidth]{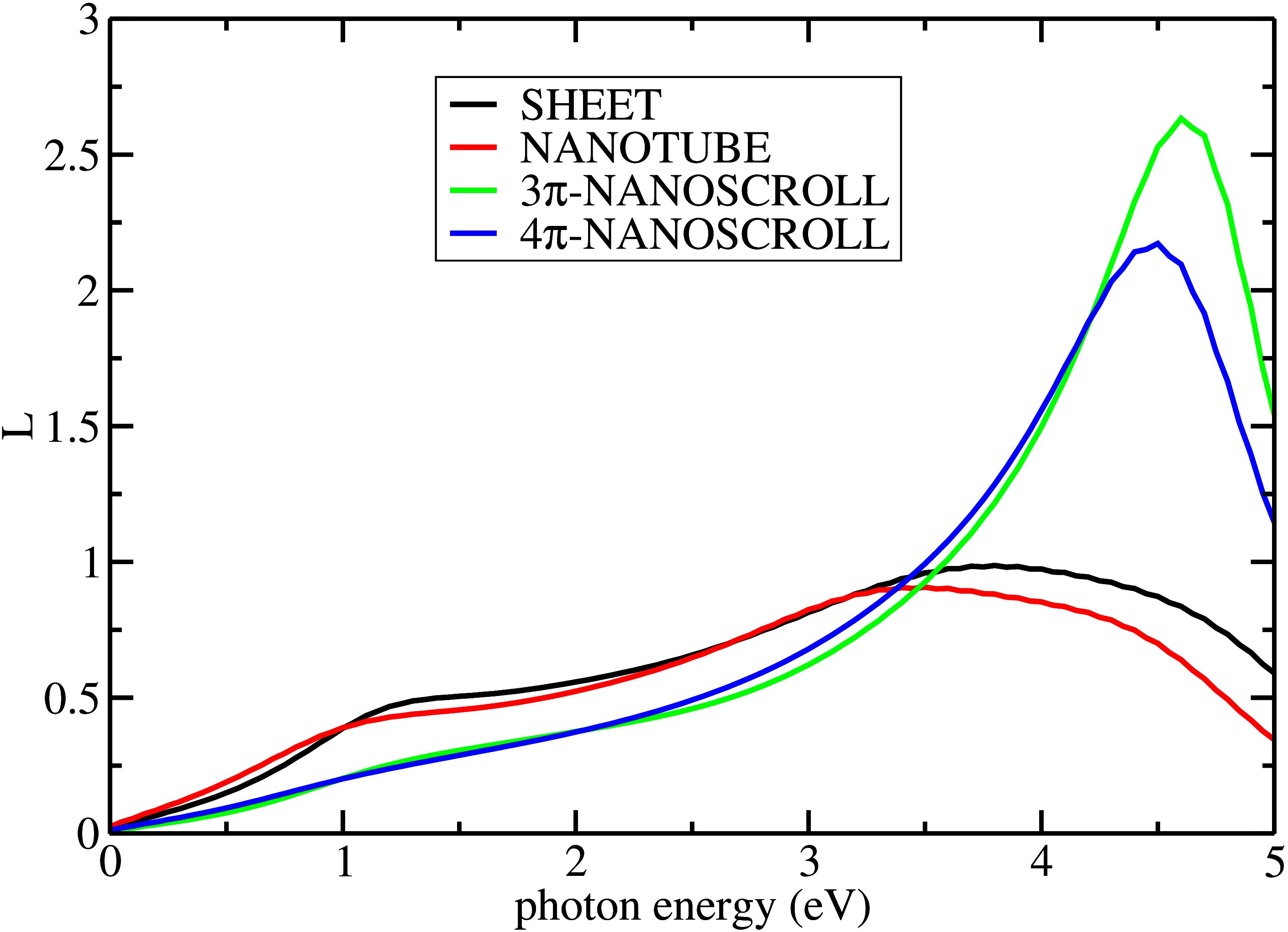}
\caption{We also calculated the loss function to determine the energy associated with plasma frequency $\omega_p$, namely plasma energy ($\hbar\omega_p$). The plasma frequency corresponds to rapid oscillations of the electron density in conducting medium within the ultraviolet region. The photon energy for the maximum peak of loss function gives us the plasma energy value. In the graphene case, the plasma energy is given by $6.05$ eV  \cite{optic1,optic2}. In Figure S\ref{fig:loss} of the Supplementary Materials we present these results. The obtained values for plasma energy: $3.9$ eV, $3.5$ eV, $4.6$ eV and $4.5$ eV for MAC-based sheet, nanotube, 3$\pi$-nanoscroll, and $4\pi$-nanoscroll, respectively.}
\label{fig:loss}
\end{center}
\end{figure}

\end{suppinfo}

\bibliography{achemso-demo}

\end{document}